\documentclass[prl,superscriptaddress, preprintnumbers,onecolumn,showpacs]{revtex4}
\usepackage{amsmath}
\usepackage{mathrsfs}
\usepackage{amssymb}
\usepackage{amsthm}
\usepackage{amsfonts}
\usepackage{algorithmic}
\usepackage{enumerate}
\usepackage{latexsym}
\usepackage{graphicx}
\usepackage{bm}
\usepackage{subfigure}
\usepackage[dvips]{color}
\usepackage{diagbox}
\usepackage[latin1]{inputenc}

\begin{document}

\title{Transport through Quantum Anomalous Hall Bilayers with Lattice Mismatch}
\author{Yan Yu}
\affiliation{SKLSM, Institute of Semiconductors, Chinese Academy of Sciences, P.O. Box 912, Beijing 100083, China}
\affiliation{School of Physical Sciences, University of Chinese Academy of Sciences, Beijing 100049, China}
\affiliation{School of Physics and Materials Science, Guangzhou University, 510006 Guangzhou, China}

\author{Yan-Yang Zhang}
\email{yanyang@gzhu.edu.cn}
\affiliation{School of Physics and Materials Science, Guangzhou University, 510006 Guangzhou, China}
\affiliation{Huangpu Research and Graduate School of Guangzhou University, 510700 Guangzhou, China}
\affiliation{School of Mathematics and Information Science, Guangzhou University, 510006 Guangzhou, China}

\author{Si-Si Wang}
\affiliation{School of Physics and Materials Science, Guangzhou University, 510006 Guangzhou, China}
\affiliation{School of Mathematics and Information Science, Guangzhou University, 510006 Guangzhou, China}

\author{Ji-Huan Guan}
\affiliation{Beijing Academy of Quantum Information Sciences, Beijing 100193, China}
	\affiliation{SKLSM, Institute of Semiconductors, Chinese Academy of Sciences, P.O. Box 912, Beijing 100083, China}

\author{Xiaotian Yang}
\affiliation{Key Laboratory of Artificial Micro- and Nano-structures of Ministry of Education and School of Physics and Technology,
Wuhan University, Wuhan 430072, China}

\author{Yang Xia}
\affiliation{Microelectronic Instrument and Equipment Research Center, Institute of Microelectronics of Chinese Academy of Sciences, Beijing 100029, China}
\affiliation{School of Microelectronics, University of Chinese Academy of Sciences, Beijing 100049,
China}

\author{Shu-Shen Li}
\affiliation{SKLSM, Institute of Semiconductors, Chinese Academy of Sciences, P.O. Box 912, Beijing 100083, China}
\affiliation{College of Materials Science and Opto-Electronic Technology, University of Chinese Academy of Sciences, Beijing 100049, China}
\affiliation{Synergetic Innovation Center of Quantum Information and Quantum Physics, University of Science and Technology of China, Hefei, Anhui 230026, China}

\date{\today}

\begin{abstract}
We theoretically investigate quantum transport properties of quantum anomalous Hall bilayers, with arbitrary ratio of lattice constants, i.e., with lattice mismatch. In the simplest case of ratio 1 (but with different model parameters in two layers), the inter-layer coupling results in resonant traversing between forward propagating waves in two layers. In the case of generic ratios, there is a quantized conductance plateau originated from two Chern numbers associated with two layers. However, the phase boundary of this quantization plateau consists of a fractal transitional region (instead of a clear transition line) of interpenetrating edge states (with quantized conductance) and bulk states (with unquantized conductance). We attribute these bulk states as mismatch induced in-gap bulk states. Different from in-gap localized states induced by random disorder, these in-gap bulk states are extended in the limit of vanishing random disorder. However, the detailed fine structure of this transitional region is sensitive to disorder, lattice structure, sample size, and even the configuration of leads connecting to it, due to the bulk and topologically trivial nature of these in-gap bulk states.
\end{abstract}

\maketitle

\section{I. INTRODUCTION}
In recent decades, topology protected states in two dimensions
attract extensive researches due to their robust quantum transports and entanglements.
Among them, the quantum anomalous Hall (QAH) effect is a two-dimensional (2D)
Chern insulator without an external magnetic field\cite{Haldane}, which possesses
quantized transport in its bulk gap, leading to potential applications in
dissipationless and quantum computing devices. After proposals based on different
materials\cite{QAH_CXLiu,QAH_HJZhang,QAH_QZWang,QAH_RYu,QAH_SCWu,QAHQSH1},
the QAH effect has been experimentally observed \cite{QAH_Exp1,QAH_Exp2,QAH_Exp3,QAH_Exp4,QAH_Exp5}.

As an effect in 2D, the minimal model of QAH effect can be realized on a monolayer
lattice with two orbitals at each site\cite{Haldane,Bernevig2006}.
On the other hand, bilayer systems have been shown to be a
platform for richer phenomena. For example, a graphene bilayer, even
regularly stacked as a periodic lattice, has possessed electronic
structures and transport properties significantly different from those of a monolayer one\cite{GrapheneRMP}.
By twisting bilayers into generic non-commensurate or quasi-periodic lattices,
more nontrivial physics emerge, including strong correlation, superconductivity
and nontrivial topology\cite{Bistritzer2011,YuanCao,XiDai2019,Twistl,Twist2}.

Quasi-periodicity is a delicate pattern between periodicity and disorder,
which may give rise to interesting behaviors of electronic wavefunctions,
even in the single particle picture\cite{FibonacciRMP,JFanReview}.
For example recently, plenty of interesting localization properties have been found
in one-dimensional (1D) lattices with a quasi-periodic potential, such as
well-defined mobility edges \cite{MobilityEdge1D1,MobilityEdge1D2}, and a critical region consisting of critical states\cite{YCZhang2022}. Insightful predictions on other rich physics in 1D quasi-periodic lattices are proposed, for example, topology, non-Hermeticity, superconductivity and superfluidity\cite{Update01,Update02,Update03,Update04}. Some interesting experimental realizations of quasicrystalline physics have also been discussed
recently\cite{HofstadterModel,Quasicrystal}.

In this manuscript, for the QAH effect, we consider another method of creating
quasi-periodicity, by coupling two QAH layers with arbitrary ratios of
lattice constants. Experimentally, this can be realized by using state-of-the-art
fabrication technologies of topological hetero-structures\cite{Fabrication1,Fabrication2,Fabrication3},
ultracold atomic\cite{FabricationOL1,FabricationOL2,FabricationOL3}
or photonic systems\cite{FabricationPhot}.
Firstly, by comparing the numerical and analytical results
in the simple case of commensurate bilayer, we confirm that the fundamental physics
underlying this system is governed by two coupled 1D Dirac equations, if no bulk states have been considered.
Then, the quantum transports are investigated for different lattice constant ratios, by using a tight-binding model that naturally includes both bulk and edge states. The central plateau of the quantized conductance is not affected since the coupling has not been able to close and re-open the bulk gap. On the other hand, the phase boundary of this quantized conductance plateau consists of a fractal transitional region, with small and fractured
islands of quantized and unquantized conductances penetrating into each other.
This picture is distinct from that of the commensurate system, where the boundaries of
the topological region are clear. Moreover, we find that these unquantized states
in the transitional region are extending across the sample. This is also distinct
from that of the disordered topological systems (topological Anderson insulators)
where they are localized. As a result, the details of this transitional region is sensitive to disorder, shape of the sample, and even the configuration of leads connected to it. The effects from nonzero disorder and varying coupling strength are also discussed.

The rest of the paper is organized as follows. In Sections II, we introduce the model and the method we use to describe the electronic transport properties of bilayer QAH system. Section III is dedicated to compare the results of commensurate lattices obtained by two different approaches: a tight-binding model using the Landauer-B\"{u}ttiker formalism and a mode-matching calculation in the continuum Dirac-like Hamiltonian approximation. Section IV is dedicated to present numerical results of the
generic case with a full tight-binding simulation: mismatching lattice constants in two layers. Finally, we conclude in Sec. V summarizing our main results.

\section{II. THE MODEL AND METHOD}

The general form of a spinless bilayer QAH system can be expressed as
\begin{equation}\label{EqFullHamiltonian}
H=H_{1}+H_{2}+H_{c},
\end{equation}
where $H_{\ell}$ ($\ell=1,2$) is the Hamiltonian for the $\ell$-th layer and
$H_{c}$ is the coupling between them. Here, we choose each layer to be the spin up component of the Bernevig-Hughes-Zhang (BHZ) model \cite{Bernevig2006} defined on a square lattice, with one $s$ orbital and one $p$ orbital on each site.
In this manuscript, we fix the lattice constant of the first layer $d_{1}=1$
as the length unit, while vary that of the second layer $d_{2}$ continuously.
The layer Hamiltonian $H_{\ell}$ in the momentum space is
\begin{equation}\label{EqBHZ1}
H_{\ell}=\sum_{\bm{k},\alpha\beta}h_{\ell;\alpha\beta}(\bm{k})c^{\dagger}_{\ell;\bm{k}\alpha},
c_{\ell;\bm{k}\beta}
\end{equation}
where $c^{\dagger}_{\ell;\bm{k}\alpha}$ ($c_{\ell;\bm{k}\beta}$) creates (annihilates) an electron with wave number $\bm{k}$ and orbital $\alpha\in{\{s,p\}}$ in the $\ell$-th layer.
For the BHZ model, $h_{\ell;\alpha\beta}$ is a $2\times2$ matrix defined as \cite{Bernevig2006}
\begin{eqnarray}\label{EqBHZ2}
h_{\ell}(\bm{k}) &=&d_{\ell}^{0}I_{2\times 2}+d_{\ell}^{1}\sigma _{x}+d_{\ell}^{2}\sigma
_{y}+d_{\ell}^{3}\sigma _{z}  \label{EqBHZ2} \\
d_{\ell}^{0}(\bm{k}) &=&-2D_{\ell}\big(2-\cos k_{x}-\cos k_{y}\big)  \notag \\
d_{\ell}^{1}(\bm{k}) &=&A_{\ell}\sin k_{x},\quad d_{\ell}^{2}(\bm{k})=A_{\ell}\sin k_{y}  \notag \\
d_{\ell}^{3}(\bm{k}) &=&M_{\ell}-2B_{\ell}\big(2-\cos k_{x}-\cos k_{y}\big),  \notag
\end{eqnarray}
with $\sigma_{x,y,z}$ the Pauli matrices acting on the orbital space $\{s,p\}$.
We assume that these model parameters of each layer can be tuned independently.
In the absence of the inter-layer coupling $H_{c}$, the $\ell$-th layer
is in the QAH state if $B_{\ell} \cdot M_{\ell}>0$\cite{QAH2010}.
The real space version of the layer Hamiltonian
$H_{\ell}=\sum_{ij,\alpha\beta}h_{\ell;\alpha\beta}(i,j)c^{\dagger}_{\ell;i\alpha},c_{\ell;j\beta}$
can be obtained from Eq.(\ref{EqBHZ1})
and (\ref{EqBHZ2}) by performing a straightforward inverse Fourier transformation $c_{\ell;\bm{k}\beta}=\frac{1}{\sqrt{V}}\sum_{j}c_{\ell;j\beta}e^{-i\bm{k}\cdot\chi_{j}}$,
where $j$ is the site index.

\begin{figure}[htbp]
	\includegraphics*[width=0.6\textwidth]{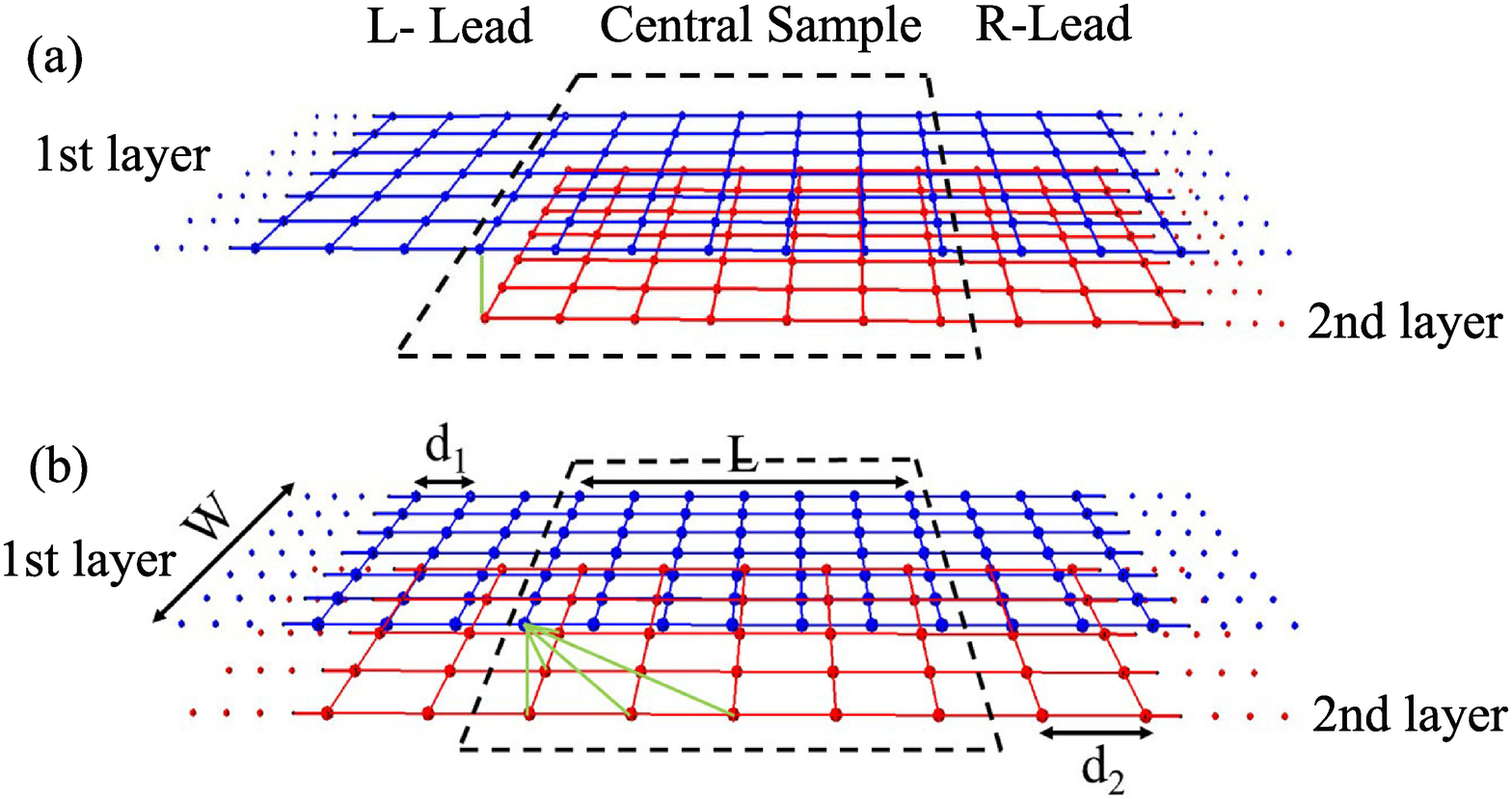}
	\caption{(Color online) Schematic of the bilayer QAH sample (enclosed by dashed-line rectangle) contacted with two semi-infinite leads. (a) The setup used in Section III, the central sample consists of two layers with identical lattice constant but different model parameters. The inter-layer bonds only connect nearest sites (green line). The left (right) lead is a QAH monolayer (decoupled bilayer) connected to the first (both) layer(s) of the central bilayer sample. (b) The setup used in remaining sections. The central sample is a bilayer with arbitrary ratio of lattice constants, with distance-dependent inter-layer couplings defined by Eq. (\ref{EqInter-layerCoupling2}) (green lines). Each lead consists of two decoupled monolayers. In both cases, the length (width) of the central bilayer sample are $L$ ($W$), in units of $d_{1}$, the lattice constant of the first layer. }
	\label{FigDevice}
\end{figure}

With the real space layer Hamiltonian at hand, we can define the inter-layer Hamiltonian $H_{c}$. In most of this manuscript (i.e., except Section III), we adopt a simple form as \cite{InterlayerCoupling1,InterlayerCoupling2}
\begin{equation}\label{EqInter-layerCoupling2}
H_{c}=\sum_{ij,\alpha}(\frac{t}{d_{ij}}c^{\dagger}_{1;i\alpha}c_{2;j\alpha}+\mathrm{H. c.}),
\end{equation}
where $t$ is the strength of the inter-layer coupling and $d_{ij}$ is the distance between sites $i$ and $j$ in the first and second layer respectively. To simplify numerical calculations, we only count in inter-layer couplings $\frac{t}{d_{ij}}\geqslant 0.01t$. By varying the lattice constant of the second layer, the lattice mismatch can be felt by the electron through $H_{c}$.

We concentrate on the transport properties of the QAH bilayer connected to two semi-infinite leads, as shown in figure  \ref{FigDevice}. At zero temperature, the two-terminal conductance can be calculated within the Landauer-B\"{u}ttiker formalism $G=\frac{2e^2}{h}T$\cite{Landauer1957,Buettiker,TransmissionRMP,Datta},
where $e$ is the elementary charge, $h$ is the Planck constant and 2 is the spin degeneracy. This transmission $T$ at Fermi energy $E$ can be expressed in terms of Green's functions as\cite{DHLee,QuantumTransport},
\begin{equation}\label{EqnTransmission}
	T(E)=\mathrm{Tr}[\Gamma_{R}G_D^{r}\Gamma_{L}G_D^{a}],
\end{equation}
where the dressed retarded (advanced) Green's function $G^{r(a)}_{D}(E)=\big[E - H -\Sigma^{r(a)}_{L}-\Sigma^{r(a)}_{R}\big]^{-1}$, $\Gamma_{p}=i\big[\Sigma^{r}_{p}-\Sigma^{a}_{p}\big]$ , and $\Sigma^{r(a)}_{p}$ is the retarded (advanced) self-energy from lead $p(= L,R)$, i.e., the left and right leads.

The local current from site $i$ to $j$ along the bond is\cite{Jiang2009}
\begin{equation}
	J_{i\rightarrow j}(E)=\frac{2e^2}{h}\mathrm{Im}\big[H_{ij}G^{n}_{ji}(E)\big](V_S-V_D),\label{EqLocalCurrent}
\end{equation}
where $H_{ij}$ the matrix element of the bare Hamiltonian $H$, and $G^{n}(E)=G^{r}_{D}(E)\Gamma_{L}(E)G^{a}_{D}(E)$ is the correlation function. Since it is defined in the linear response regime, we simply take the voltage difference $V_S-V_D$ between the source and drain leads to be unity.

To understand more physics underlying transport properties, we rely on the
density of states (DOS) \cite{YYZhang2013} and the normalized
participation ratio (NPR) \cite{NPR2021,NPR2020,NPR2017}.
With the real space representation of the Green's function $G^{r}(E;i,j)$,
the single particle local density of states (LDOS) $\rho(i)$ at each site $i$,
and the total density of states (DOS) can be calculated respectively as
\begin{eqnarray}
    \rho(E,i)&=&-\mathrm{Im}G^{r}(E;i,i)\\ \label{LDOS1}
	\mathrm{DOS}(E)&\equiv &\frac{1}{N}\sum_{i=1}^{N}\rho(i) \label{LDOS1}
\end{eqnarray}
where $N$ is the total number of sites, and the summation is over all sites of the sample. Here the Green's function $G^{r}$ can be the dressed one $G^{r(a)}_{D}(E)$ defined above,
or the bare one $G^{r(a)}_{B}(E)=\big[(E \pm i\eta)I - H \big]^{-1}$,
depending on whether one wants to include the effects from leads.
With the LDOS normalized over the sample $\tilde{\rho}(i)$,
the NPR can be defined as follows\cite{NPR2021,NPR2020,NPR2017}
\begin{eqnarray}
    \tilde{\rho}(i)&=&\frac{\rho(i)}{\sum_{i=1}^{N}\rho(i)},\\ \label{EqNPR1}
	\mathrm{NPR}&=&(N\sum_{i=1}^{N}\tilde{\rho}(i)^{2})^{-1}, \label{EqNPR2}
\end{eqnarray}
The NPR is a significant diagnostic tool to characterize the localization transition. The localized (extended) phases are characterized by $\mathrm{NPR}=0$ ($\mathrm{NPR}\neq0$) in the large $N$ limit \cite{NPR2021,NPR2020,NPR2017}.

To visualize and distinguish edge states from bulk ones, the edge-locality marker is introduced as\cite{HofstadterInsulators}
\begin{equation}\label{EdgelocalityMarker}
	\mathcal{B}=\sum_{i\in \mathrm{edge}}\tilde{\rho}(i),
\end{equation}
where the summation runs over four edges of the bilayer QAH sample.
In this work, we take the width of the edge as 10\% of that of the bilayer sample.

\section{III. THE COMMENSURATE BILAYER SYSTEM}

In the topological phase of the bilayer, although the backscattering is forbidden, the inter-layer (forward) scattering is not. The nontrivial topology only guarantees the robustness of the \emph{total} transmission through the bilayer, but the details of inter-layer scattering are not clear yet.
As a starting point, let us gain some physical insights from the simplest bilayer with identical lattice constants but different model parameters. To further simplify the model so that an analytical continuum model can be easily obtained,
we adopt a simpler form of inter-layer coupling in this section as
\begin{equation}\label{EqInter-layerCoupling1}
	H_{c}=\sum_{i\alpha}(tc^{\dagger}_{1;i\alpha}c_{2;i\alpha}+\mathrm{H. c.}),
\end{equation}
where the inter-layer coupling $t$ [green line in figure  \ref{FigDevice}(a)] only exists between nearest sites from different layers.

\begin{figure}[htbp]
\includegraphics*[width=0.45\textwidth]{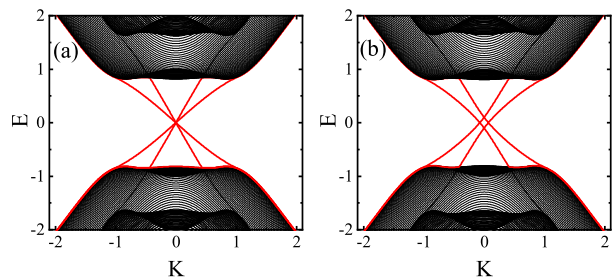}
\caption{(Color online) The band structure of the bilayer QAH ribbon with identical lattice constants, and with the width $W = 100$. (a) $t = 0$. (b) $t=0.1$. The rest model parameters are identical for both panels: $A_{1} = 1$, $B_{1} = -1$, $D_{1} = 0$, $M_{1} = -1$, $A_{2} = 2$, $B_{2} = -2$, $D_{2} = 0$, and $M_{2} = -2$. The red curves are edge states. }
\label{FigBandStructure}
\end{figure}

In figure  \ref{FigBandStructure}, we present the band structures of the QAH bilayer in the ribbon geometry, without (a) and with (b) the inter-layer coupling respectively. With remarkably different model parameters in two layers, the velocities of two groups of edge states are different too.
In figure  \ref{FigBandStructure}(b), the inter-layer coupling lifts the trivial degeneracy of edge states at the $\Gamma$ point.

For this commensurate system with the inter-layer coupling (\ref{EqInter-layerCoupling1}), the low-energy properties of the QAH edge states can be captured by a Dirac-like continuum model from $k \cdot p$ approximation \cite{Dirac-likeHamiltonian1,Dirac-likeHamiltonian2}.
This effective Hamiltonian has the form\cite{BilayerHamiltonian1,BilayerHamiltonian2}:
\begin{equation}\label{DiracLikeHam}
	\mathscr{H}=\left(
	\begin{array}{cc}
		v_{1}k_{x}   & F      \\
		F^{\dag} & v_{2}k_{x} \\
	\end{array}
	\right),
\end{equation}
where $k_{x}$ is the $x$ component of the momentum relative to the Dirac point, and $v_{i}>0$ is the velocity of edge states in the $i$-th layer. For  each layer, only the forward channel should be considered here, due to the forbidden of backscattering. The inter-layer coupling coefficient $F$ depends on the concrete realization of the bilayer. It can be verified that, when the model parameters of the bilayer QAH system satisfy $\frac{A_{2}}{A_{1}}=\frac{B_{2}}{B_{1}}=\frac{D_{2}}{D_{1}}=\frac{M_{2}}{M_{1}}$, $F$ just equals the coupling strength $t$ in Eq. (\ref{EqInter-layerCoupling1}). In the rest of this section, we adopt the model parameters satisfying this condition. In the low energy limit around $k_{x} = 0$, we have $v_{\ell}=\mathrm{sgn}(B_{\ell})A_{\ell}\sqrt{1-\frac{D_{\ell}^{2}}{B_{\ell}^{2}}},\ell=1,2$ \cite{Dirac-likeHamiltonian1,Dirac-likeHamiltonian2}. The expressions of eigenvalues and eigenfunctions of this coupled Dirac Hamiltonian are listed in the Appendix.

\begin{figure}[t]
\includegraphics*[width=0.7\textwidth]{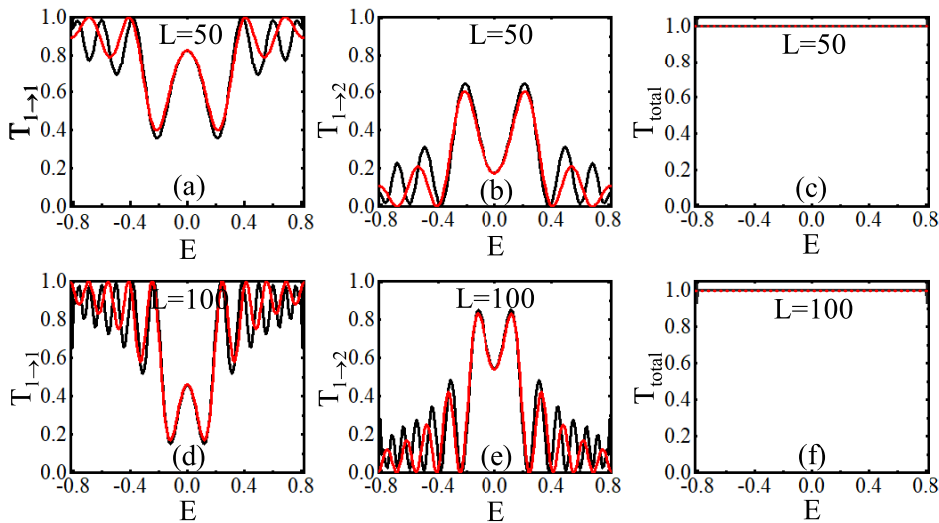}
\caption{(Color online) Transmissions as functions of Fermi energy with length $L = 50$ [first row] and $L = 100$ [second row]: the intra-layer transmission $T_{1\rightarrow 1}$ (left column), the inter-layer transmission $T_{1\rightarrow 2}$ (middle column), and the total transmission $T_{\mathrm{total}}$ (right column). Black (red) lines are results from the lattice model (effective continuum model). Other model parameters are identical for both rows: $A_{1} = 1$, $B_{1} = -1$, $D_{1} = 0$, $M_{1} = -1$, $A_{2} = 2$, $B_{2} = -2$, $D_{2} = 0$, $M_{2} = -2$, $t = F = 0.1$ and $W = 60$.}
\label{EnergyConductance}
\end{figure}

\begin{figure}[t]
\includegraphics*[width=0.45\textwidth]{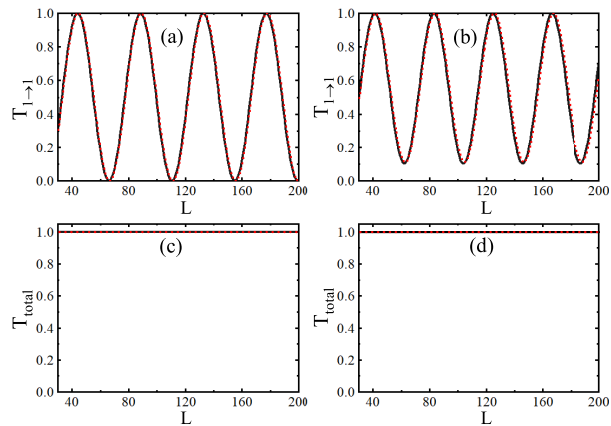}
\caption{(Color online) Transmissions as functions of the central sample length $L$ at a Fermi energy $E = 0$ (left column), and $E = 0.1$ (right column). The first (second) row corresponds to the intra-layer transmission $T_{1\rightarrow1}$ (the total transmission $T_{\mathrm{total}}$). Black solid (red dashed) lines are results from the lattice model (effective continuum model). The rest model parameters are: $A_{1} = 1$, $B_{1} = -1$, $D_{1} = 0$, $M_{1} = -1$, $A_{2} = 2$, $B_{2} = -2$, $D_{2} = 0$, $M_{2} = -2$, $t = F = 0.1$ and $W = 60$.}
\label{LConductance}
\end{figure}

We use the transport configuration illustrated in figure  \ref{FigDevice}(a), with one layer as the source lead while a \emph{decoupled} bilayer as the drain lead, so that the intra- and inter-layer transports can be conveniently distinguished. Since the source lead is a monolayer carrying Chern number 1, the total transmission through the central sample will be quantized as 1, If it is in the topological state.
Then due to the inter-layer coupling of the sample, an electron from the source lead of monolayer can be transmitted into either monolayer of the drain lead, corresponding to the intra-layer and inter-layer transmissions respectively.
It is straightforward to solve this tunnelling problem.
Due to one monolayer (with one pair of edge states) as the source,
only one active channel (positive-velocity branch of this pair) is propagating in the central QAH bilayer, with the length $L$ and width $W$. The wave functions in different spatial regions can be written as
\begin{eqnarray}
\psi(x)= \begin{cases}
  \left(
  \begin{array}{c}
              e^{ik_{1}x}\\
              0          \\
  \end{array}
  \right),&x<0 \\
  M\psi_{1}(x)+N\psi_{2}(x),&0<x<L \\
  t_{1}\left(
  \begin{array}{c}
              e^{ik_{1}x}\\
              0          \\
  \end{array}
  \right)+t_{2}\left(
  \begin{array}{c}
              0          \\
              e^{ik_{2}x}\\
  \end{array}
  \right),&x>L \\
\end{cases}
\end{eqnarray}
where $k_{\ell}=\frac{E}{v_{\ell}},\ell=1,2$ at the incoming energy $E$.
In the central bilayer sample the wave functions are linear combinations of the eigenfunctions given in Eq. (\ref{Eigenvectors}) in Appendix, with $M$ and $N$ the combination coefficients. Due to the perfect transmission of Dirac-like fermions, there is no reflection here\cite{PerfectTransport1,PerfectTransport2,PerfectTransport3}.
We only need to match the wave functions at boundaries because the Dirac-like
Hamiltonian is first-order differential\cite{KleinParadox,GrapheneProperties}. From this continuity of the wave functions at the $x=0$ and $x=L$, i.e., boundaries of the central bilayer, we obtain coefficients $M$, $N$, $t_{1}$ and $t_{2}$. According to current conservation of the Dirac-like Hamiltonian,
we have $T_{1\rightarrow1}=t_{1}t_{1}^{\ast},T_{1\rightarrow2}=\frac{v_{1}t_{2}t_{2}^{\ast}}{v_{2}}$\cite{Current1,Current2,Current3,Current4}, where $T_{1\rightarrow1}$ ($T_{1\rightarrow2}$) represents the intra-layer (inter-layer) transmission.

In figure \ref{EnergyConductance}, red curves are results calculated from this effective continuum model, for the intra-layer (left column)
and inter-layer (right column) transmissions, with the central sample length
$L=50$ (upper row) and $L=100$ (lower row) respectively.
For comparison, corresponding results from the tight-binding simulations are also plotted as the black curves. The perfect quantization of the total transmission shown in figure \ref{EnergyConductance} (c) and (f) confirm that the central sample is in the topological phase.

Now let us scrutinize details of the layer resolved transmissions $T_{1\rightarrow 1}$ and $T_{1\rightarrow 2}$. Curves in both colors agree very well, especially when $E\rightarrow0$. So in the low energy limit, the Dirac continuum approximation contains most of the physics. The intra-layer and inter-layer transmissions show a complementarity: the transmission minima in one configuration coincide with the maxima of the other one, since the sum of them is robustly quantized as 1.

Using the above mentioned analytical formalism, we can also obtain the dependence on the sample length $L$ as
\begin{equation}\label{ZeroTransmission}
T_{1\rightarrow1}=\frac{\cos(2q_{1}L)+1}{2}\ \mathrm{when}\ E=0,
\end{equation}
where $q_{1}=\frac{F}{\sqrt{v_{1}v_{2}}}$ by Eq. (\ref{Eigenvectors}).
In figure \ref{LConductance}, we can clearly see that for a fixed energy $E$, the transmission varies periodically with $L$ from . At zero energy. the period $\frac{\pi\sqrt{v_{1}v_{2}}}{F}$ can be obtained from Eq. (\ref{ZeroTransmission}), which has an excellent agreement with the tight-binding result shown in figure \ref{LConductance}(a).

\begin{figure}[htbp]
	\centering
	\includegraphics*[width=0.5\textwidth]{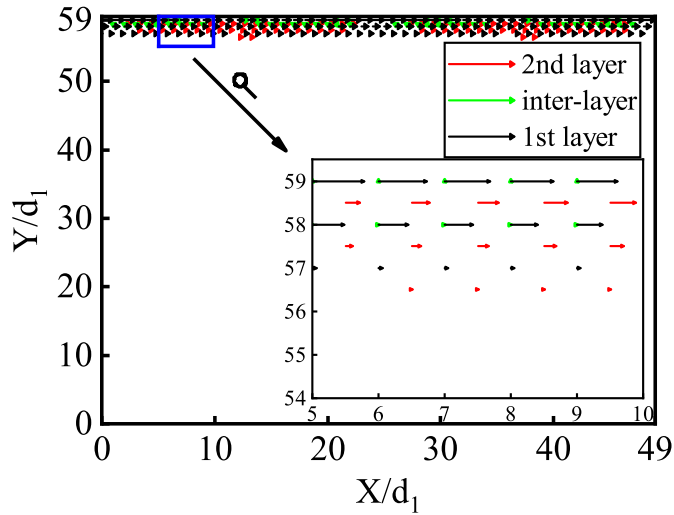}
	\centering
	\caption{ Configurations of local currents in the central sample with length $L=50$ at Fermi energy $E=0.39$, with the arrow size proportional to the magnitude of the current. The inset is enlargement of the region marked by the blue-line rectangle. Black (red) arrows correspond to currents in the first (second) layer, and green arrows the inter-layer currents. For a better visual clarity, the second layer has been shifted diagonally by $\frac{\sqrt{2}}{2}d_1$. The rest model parameters are identical to figure  \ref{EnergyConductance}(a).}
	\label{FigLocalCurrent}
\end{figure}

In Fig. \ref{FigLocalCurrent}, we plot the spatial distribution of local currents [obtained from equation (\ref{EqLocalCurrent})] of the sample at a typical parameter setting (see the caption), which clearly show that the currents are carried by edge states.

In brief, in the topological phase, if the influence from bulk states are ignored, the transmission through the bilayer is quantized. Now the only interesting effects from the inter-layer coupling are just the resonant traversing between forward propagating waves in two layers.

\section{IV. Generic Ratios of Lattice Constants between Bilayers}

In the rest of this manuscript, we will come back to the generic case of
mismatching lattice constants between two layers, with the distance dependent inter-layer coupling Eq. (\ref{EqInter-layerCoupling2}). As mentioned in Section II, the lattice constant of the first layer $d_{1}$ is fixed to be the length unit, and that of the second layer $d_{2}$ will be varied. Now even in the topological phase, a low energy approximation based analytical treatment will be difficult, because, for example,
the inter-layer coefficient $F$ in the effective Hamiltonian (\ref{DiracLikeHam})
will have a very complicated dependence on model parameters and lattice configurations.
Moreover, the effective Dirac Hamiltonian (\ref{DiracLikeHam}) only involves edge states and
therefore it cannot describe any effects from bulk states, or the transition into topologically trivial phase, which we are focusing in the following.
As a result, we will rely on numerical calculations based on the tight-binding model, as introduced in Section II.

\begin{figure*}[htbp]
\includegraphics*[width=0.9\textwidth]{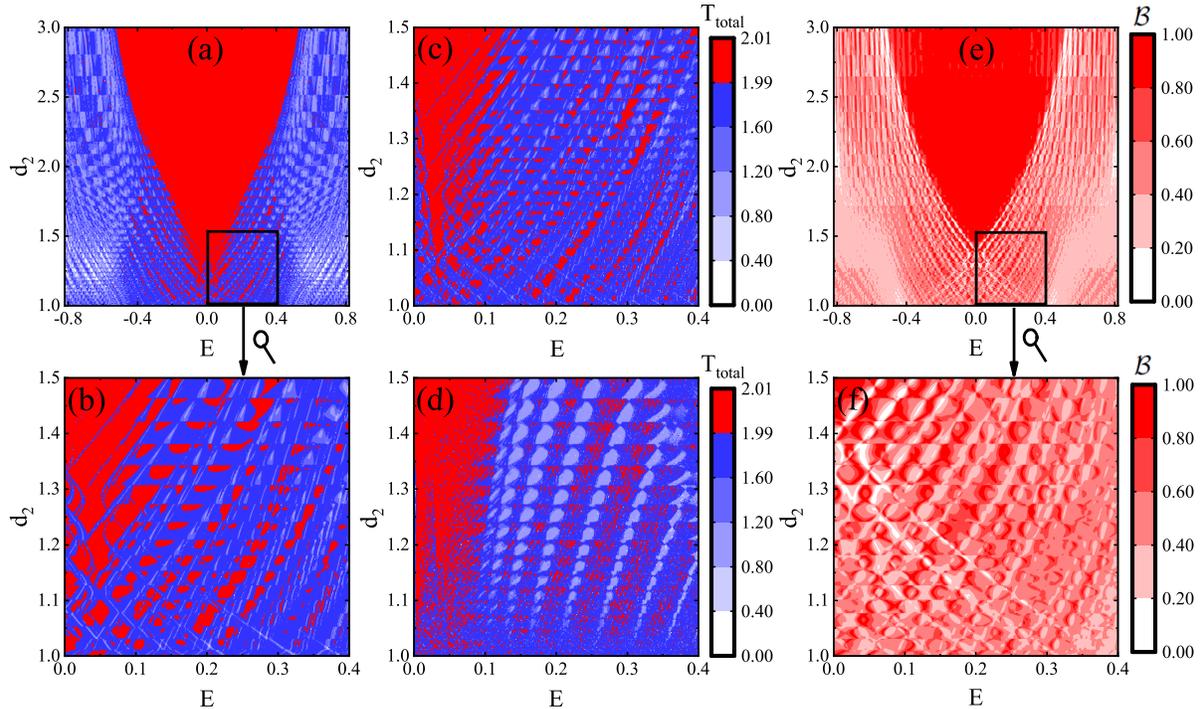}
\caption{(Color online) Transport phase diagram on the parameter plane of $d_{2}-E$. (a) The total transmission $T_{\mathrm{total}}$ for the central bilayer QAH sample with length $L = 50$ and width $W=50$. (b) Enlargement of the region enclosed by the black-line square in figure  \ref{ConductanceEdgelocalityMarker}(a). In this same enlarged region, the total transmission for the central bilayer QAH sample with (c) $L=W=80$ and (d) $L=500$, $W=50$. (e) The edge-locality marker $\mathcal{B}$ associated with panel (a). (f) Enlargement of the region enclosed by the black-line square in panel (e), or, $\mathcal{B}$ associated with panel (b). The rest model parameters are: $A_{1} = 1$, $B_{1} = -1$, $D_{1} = 0$, $M_{1} = -1$, $A_{2} = 1$, $B_{2} = -1$, $D_{2} = 0$, $M_{2} = -1$ and $t=0.1$.}
\label{ConductanceEdgelocalityMarker}
\end{figure*}

The transport configuration will be taken to be a more ``natural'' setup as illustrated in figure  \ref{FigDevice}(b).
Each lead (source or drain) consists of two \emph{decoupled} monolayers,
otherwise the coupling between incommensurate bilayers will break the
translation symmetry of the lead and make the self energies incomputable.
Since we are not interested in what happens in the leads,
this setup will not influence the main physics of the results.
The model parameters of each monolayer lead are identical to those of the central sample layer it is connected to, to decrease unwanted scattering on the contact boundaries.
The total transmission is a sum of all four possible transmissions between these monolayer leads, i.e., $T_{\rm{total}} = T_{1\rightarrow1}+T_{1\rightarrow2}+T_{2\rightarrow1}+T_{2\rightarrow2}$.

\subsection{A. Phase Diagram on the $E-d_2$ Plane}

Let us consider a system with $A_{1}=A_{2}=1, B_{1}=B_{2}=-1, D_{1}=D_{2}=0, M_{1}=M_{2}=-1$, where each monolayer is topologically nontrivial with Chern number 1. The topological phase of the bilayer can be identified as the quantized transmission $T_{\rm{total}} = 2$ with the contribution from edge states in both layers. When the lattice constant of the second layer $d_{2}$ is varied gradually, the bilayer system displays interesting transport properties in the presence of inter-layer coupling. figure  \ref{ConductanceEdgelocalityMarker}(a)-(d) shows the phase diagram of the total transmission on the $d_{2}-E$ plane with the open boundary condition. In figure  \ref{ConductanceEdgelocalityMarker}(a), the most prominent feature is the central plateau of the quantized conductance (red).
The width of this central plateau grows larger with increasing $d_2$.
This can be understood as follows. With the stretching of the second layer,
most of the inter-layer bonds becomes longer and therefore the associated coupling $t$ becomes smaller. In other words, the effective ``average'' coupling between two layers becomes weaker. As a result, the bilayer somehow tends to approach the decoupling case, with a large bulk gap around $E=0$ where the edge states live.

More interesting behaviors emerge on the boundary region of this central plateau of quantization. In the parameter space of a periodic lattice, the phase boundary of the topological state is a clear-cut line\cite{Bernevig2006}.
Here instead, as shown in figure  \ref{ConductanceEdgelocalityMarker}(a) and its partial enlargement (b), the boundary consists of a fractal transitional region with interpenetrating quantized (red) and unquantized (blue) states, especially near the lower tip of the central plateau with $d_2\sim 1$. In this same enlarged region, figure  \ref{ConductanceEdgelocalityMarker}(c) and (d) show results for different sample sizes. Although unquantized states (blue) flood in, a large number of small islands of quantization with $T=2$ (red) still survive in a regular pattern of distribution. With increasing size [figure  \ref{ConductanceEdgelocalityMarker}(c) and (d)], this picture of floods and islands will be more fragmented, with quantized and unquantized parts penetrating into each other with more fine details.

\begin{table}[htbp]
	\caption{The box counting dimension of the curve $T_{\mathrm{total}}(E)$ for different sample sizes and different $d_{2}$.}
	\label{tab:sample}
	\centering
	\footnotesize
	\setlength{\tabcolsep}{4pt}
	\renewcommand{\arraystretch}{1.2}
	\begin{tabular}{|c|c|c|c|}
		\hline
        &$L=W=50$&$L=W=80$&$L=500, W=50$\\
		\hline
		$d_2=$1.2&1.35&1.47&1.55\\
		\hline
		$d_2=$1.3&1.32&1.43&1.50\\
		\hline
		$d_2=$1.4&1.31&1.35&1.51\\
		\hline
		$d_2=\sqrt{2}$&1.26&1.41&1.45\\
		\hline
	\end{tabular}
\end{table}

To shed further light on the fractal nature of the transitional region, we calculate the dimension of the curve $T_{\mathrm{total}}(E)$ by using a box-counting (BC) algorithm \cite{box-counting1}. This algorithm counts the number $N$ of squares of size $\delta$ which are necessary to continuously cover the graph of $T(E)$ rescaled to a unit square. We choose an intermediate region (usually called the ``scaling region'') where scaling is linear in a log-log plot, i.e., where $N \sim \delta^{-d} $. The slope $d$ in the scaling region is the estimate of the BC dimension \cite{box-counting2}. In Table 1, we show the results of the BC dimension for different sample sizes and different $d_{2}$. With increasing size, the BC dimensions get larger, which is consistent with above knowledge obtained from the figure \ref{ConductanceEdgelocalityMarker} that more finer details emerge.

\begin{figure}[htbp]
\includegraphics*[width=0.4\textwidth]{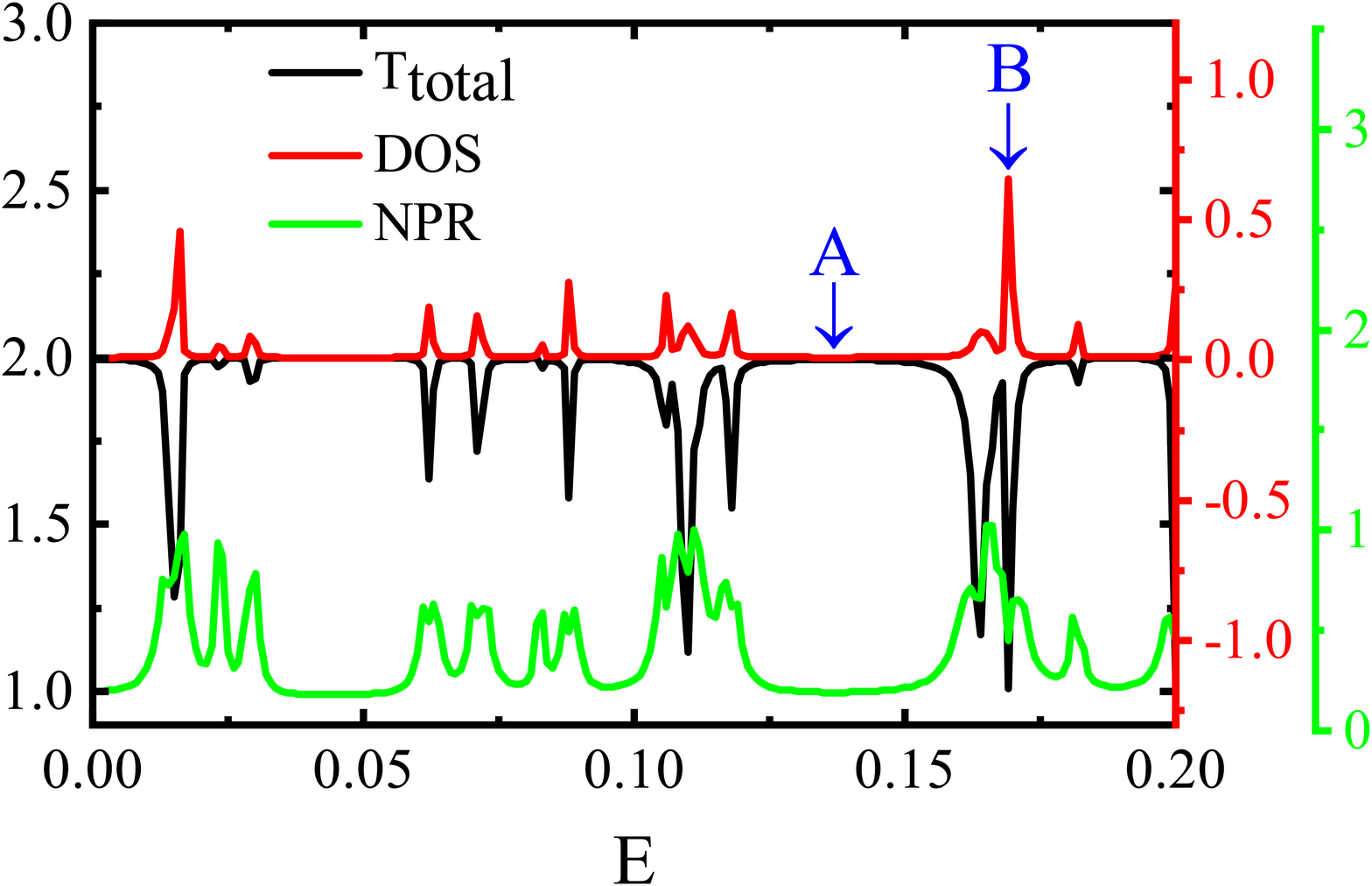}
\caption{(Color online) The total transmission (black solid lines), DOS (red solid lines) and NPR (green solid lines) as a function of Fermi energy $E$ for a fixed $d_{2}=1.2$. Other model parameters are identical to figure  \ref{ConductanceEdgelocalityMarker}(a). The real space distribution of LDOS associated with point A (B) will be plotted in the first (second) row of figure  \ref{LDOSdistribution}.}
\label{DOSConductanceNPR}
\end{figure}

Now we will investigate the property and origin of this picture of transports. Let us first scrutinize this by relating transports with electronic wavefunction behaviors. In figure  \ref{ConductanceEdgelocalityMarker}(e), the edge-locality marker $\mathcal{B}$ defined in Eq. (\ref{EdgelocalityMarker}) is plotted on the same $E-d_2$ plane, and also with figure  \ref{ConductanceEdgelocalityMarker}(f) its partial enlargement.
Here the LDOS was calculated from an isolated bilayer sample (i.e., without being connected to leads) in order to manifest the intrinsic property of the bilayer structure itself. It is interesting to notice the perfect correspondence between $T$ and $\mathcal{B}$ [(a) and (e); (b) and (f)]: a quantized $T=2$ corresponds to an edge state with $\mathcal{B}\sim 1$, while an unquantized $T$ corresponds to a bulk states with $\mathcal{B}\ll 1$. In other words, the phase boundaries of the central plateau of conductance quantization consist of a transitional region, where a series of inter-crossing bulk states are imbedded in to separate the topological edge states into small islands.

To disclose more details of this picture, we plot two partly cross sections of
figure  \ref{ConductanceEdgelocalityMarker}(b) along $E$ as the black curve in figure  \ref{DOSConductanceNPR}. The DOS and NPR are also plotted as the red and green curves respectively.
Let us first compare the transmission (black) and the DOS (red). The transmission plateau with $T=2$ always corresponds to a vanishing DOS, suggesting an edge state in the bulk gap. The peaks of the DOS correspond to trivial bulk states, which disrupt the transmission quantization.
As for the NPR (green curve), that of the trivial bulk states is always larger than that of the edge states. This suggests that these trivial bulk states are quite extended. In figure  \ref{DOSConductanceNPR}, point A (B) represents a typical localized edge (extended bulk) state.

\begin{figure}[htbp]
\includegraphics*[width=0.55\textwidth]{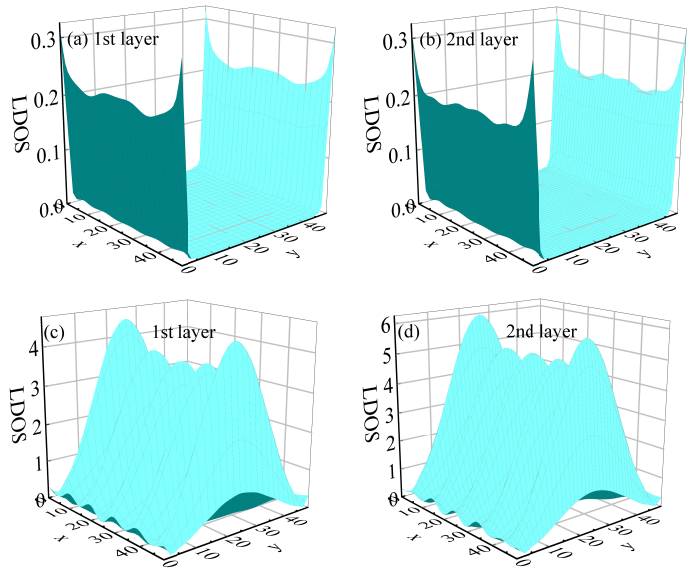}
\caption{(Color online) Real space distributions of LDOS on two layers. The first (second) row corresponds to the energy point A (B) indicated in figure  \ref{DOSConductanceNPR}. The left (right) column corresponds to the first (second) layer of the bilayer sample. }
\label{LDOSdistribution}
\end{figure}

In order to have a more direct picture of these states, we present the distributions of LDOS in figure  \ref{LDOSdistribution}. The upper (lower) row corresponds to the energy point labeled as A (B) in figure  \ref{DOSConductanceNPR}. At point A, the wavefunctions are well localized at the edges. At point B, on the other hand,
the wavefunctions are distributed in a well extended way throughout the sample bulk. This is consistent with above knowledge obtained from the NPR. As we have seen from figure  \ref{DOSConductanceNPR} and figure  \ref{LDOSdistribution}, although these bulk states are narrow, they are quite extended. This is different from those extremely narrow and localized states in the bulk of the topological Anderson insulator\cite{TAI,YYZhang2013}.
The concrete configuration (positions, widths) of these bulk states depends on the details of the lattice sample, e.g., the sample size and lattice constant ratio $d_2/d_1$. Moreover, their transports are even sensitive to the device details, as will be presented in the following.

\begin{figure}[htbp]
\includegraphics*[width=0.7\textwidth]{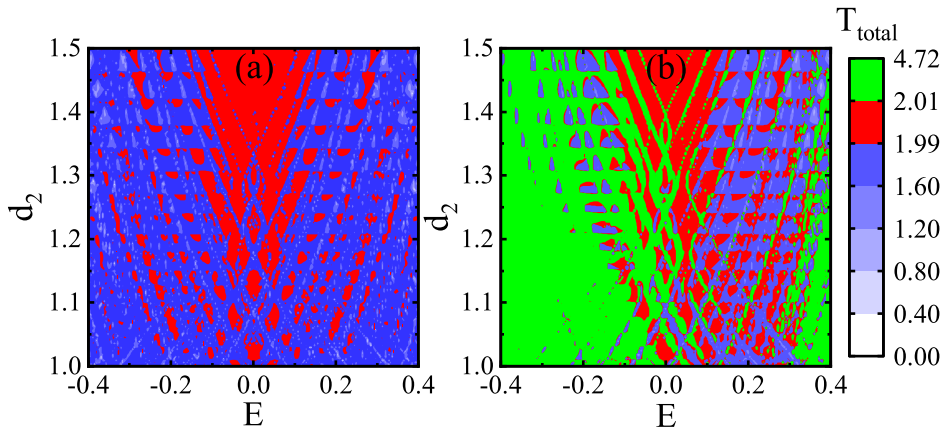}
\caption{(Color online) (a) Enlargement of the middle lower part of figure  \ref{ConductanceEdgelocalityMarker}(a). (b) Similar to panel (a) but with a higher chemical potential $\mu=1$ in leads, so that they can supply more active channels.  }
\label{FixIncidentEnergy}
\end{figure}

So far in the transport calculations, the chemical potentials of leads $\mu$ are set to vary with that in the bilayer sample. In the energy region we have focused, they are in the bulk gap of leads, so that each lead just contributes two robust channel of topological edge states. Now in order to check the robustness of the transitional region, we try to supply more active channels into the bilayer sample. In the calculation, this can be simply realized by changing the chemical potentials $\mu$ of the leads into their bulk band. The transmission
with this setup on the $E-d_2$ parameter plane is plotted in figure  \ref{FixIncidentEnergy}(b). For comparison, the same range of figure  \ref{ConductanceEdgelocalityMarker}(a) (with lead chemical potentials in the bulk gap) is re-plotted here as figure  \ref{FixIncidentEnergy}(a). In figure  \ref{FixIncidentEnergy}(b), although the profile of those crossing bulk states can still be seen, the transmission is very asymmetric respect to $E=0$. In the negative energy region, most of the quantization islands with $T=2$ have been submerged by bulk states with $T\neq2$. This can be attributed to the coexistence of edge states and in-gap bulk states. It is known that the coupling to bulk states can severely destroy the quantization and robustness of topological edge states\cite{YYZhang2014}. Furthermore, the transports of bulk states depend sensitively on details of the device due to their spatial extensions over the bulk. Recently, similar quantum transport phase transition from varying lead chemical potential is found in one-dimensional long range systems\cite{MeasurementPhaseTransition}.

\begin{figure}[htbp]
\includegraphics*[width=0.35\textwidth]{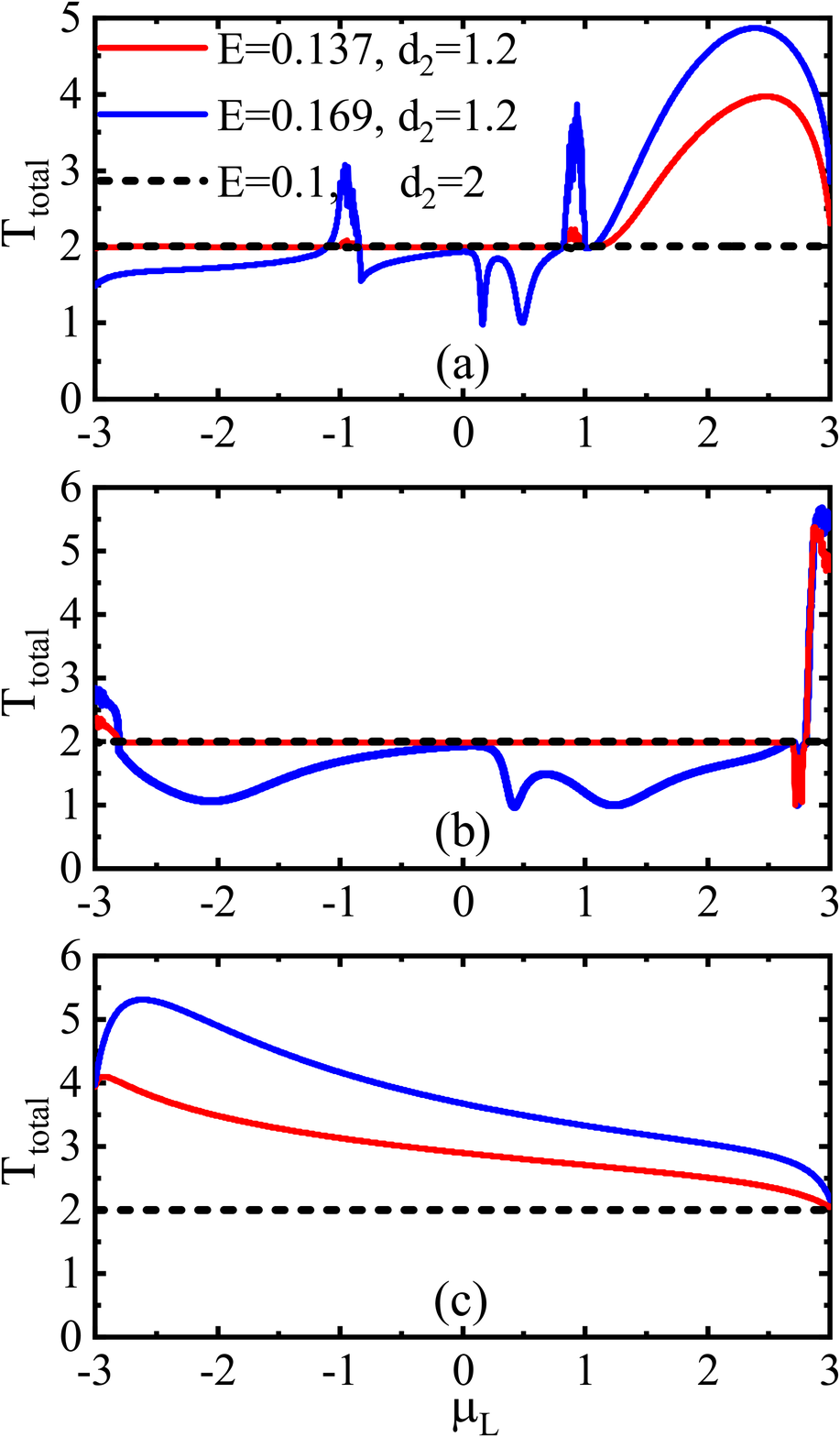}
\caption{(Color online) Total transmission as a function of the Fermi energy of the lead $\mu$ in different lead setups, with the Fermi energy of the central sample $E_F$ fixed.  Curves in different colors are results from different $E_F$ and $d_2$ fixed.  (a) Except $\mu$, other model parameters $A_{\ell}$, $B_{\ell}$, $C_{\ell}$, $D_{\ell}$ and $M_{\ell}$, are identical to the central sample. (b) Model parameters in the lead are significantly different from those in the central sample as: $A_1=A_2=3.496$, $B_1=B_2=-3.702$, $M_1=M_2=-3.5$, and $C_{\ell}=D_{\ell}=0$. (c) Leads consist of semi-infinite and decoupled 1D chains with nearest hopping $1.52$. }
\label{FigDifferentLead}
\end{figure}

To see impacts from leads more clearly, now we present the total transmission as a function of the lead Fermi energy $\mu$ (with respect to its half filling level), with all parameters of the central bilayer fixed. Three panels are for three different lead setups in figure \ref{FigDifferentLead}. Three colors of curves correspond to three typical settings of the central bilayer sample shown in figure  \ref{ConductanceEdgelocalityMarker}(a-d): on the central red plateau of quantization, on a red island of quantization, and in the blue sea of unquantization, respectively. In figure \ref{FigDifferentLead} (a), model parameters $A_{\ell}$, $B_{\ell}$, $C_{\ell}$, $D_{\ell}$ and $M_{\ell}$ are same for the sample and the leads. For a central bilayer sample on the central plateau of quantization (black dashed curve), the transmission is robustly quantized for all $\mu$, suggesting an edge state deep inside the bulk gap. For a central bilayer sample on a red island of quantization (red curve), the transmission is only quantized until $\mu\sim 1$. After that, it rises up beyond 2, showing a mixing from bulk states. For the last case, a central bilayer sample in the blue sea of unquantization (blue curve), the transmission is not quantized at all, exhibiting typical resonance peaks and valleys of bulk states.

Similar phenomena (perfect quantization on the central island of quantization, partially quantization in the transitional region, and unquantization elsewhere) can be observed from figure \ref{FigDifferentLead} (b), with significantly different model parameter settings in the lead, as shown in the figure caption.
Furthermore we use another completely different type of leads consisting of 1D chains with nearest hopping 1.52 and the results are presented in figure \ref{FigDifferentLead} (c). Now only the robust central quantization plateau (black curve) survive, and the transitional region is completely unquantized due to the strong interruptions from leads with a trivial topology.

\begin{figure}[htbp]
\includegraphics*[width=0.6\textwidth]{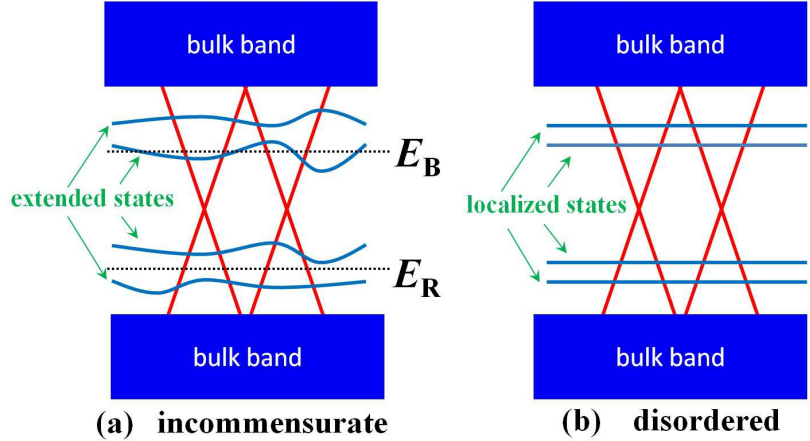}
\caption{(Color online) Schematic of electronic structure for QAH bilayers, (a) with incommensurateness from lattice mismatch, and (b) with random disorder. Red lines represent topological edge states, and blue lines represent in-gap states.
In panel (a), the Fermi energy $E_{\mathrm{B}}$ ($E_{\mathrm{R}}$) corresponds to a point in the blue sea (on the red island) of figure  \ref{ConductanceEdgelocalityMarker}(a)-(d). In both panels the Bloch theorem is broken, and therefore the horizontal axis does not necessarily mean the conventional wavevector, but other model parameters (see main text). }
\label{FigIllustration}
\end{figure}

The above results lead to a physical picture of incommensurate QAH bilayers briefly illustrated in figure  \ref{FigIllustration}(a). Here blue rectangles represent bulk bands and red curves are topological edge states between them. Since the incommensurateness breaks the standard Bloch theorem, here the horizontal axis does not necessarily stand for the conventional wavevector. Instead it can represent any other variable model parameters, e.g., lattice mismatch $d_2/d_1$. Our above results conclude that incommensurateness between two layers gives rise to a cluster of bulk states (blue curves) penetrating into margins of the bulk gap. Here we intentionally plot them as tortuous curves to stress that their positions and energy widths may be sensitive to model parameters (including coupling of leads as observed just now in figures \ref{FixIncidentEnergy} and \ref{FigDifferentLead}). These incommensurateness induced bulk states are narrow (but not flat), topologically trivial and spatially extended. We call them in-gap extended states (IGESs). Around the bulk gap center, there are only edge states (red lines), which contribute to a very robust transport, giving rise to the central red plateau of quantization in figure  \ref{ConductanceEdgelocalityMarker}(a). In the marginal region, the existence of IGES makes things complicated. If the Fermi energy $E_{\mathrm{R}}$ happens to be located in a sub-gap between IGESs, the quantization will survive, which results in a small red island of quantization in figure  \ref{ConductanceEdgelocalityMarker}(a)-(d). Otherwise, on another Fermi energy $E_{\mathrm{B}}$ with a coexistence of edge states and bulk IGESs, the quantization will be destroyed\cite{YYZhang2014}. This corresponds to blue regions in \ref{ConductanceEdgelocalityMarker} (a)-(d). Due to their bulk and topologically trivial nature, the concrete configurations (e.g., spatial distribution and width along energy axis) of IGESs will depend sensitively on details of the sample and device, as we have seen in figure  \ref{FixIncidentEnergy}.
The parameter region of IGES corresponds to the transitional region outside the central quantization plateau in figure  \ref{ConductanceEdgelocalityMarker}(a)-(d).

For comparison, the picture of purely disordered QAH effect is also shown as figure  \ref{FigIllustration}(b). Here, disorder induces in-gap localized states (blue horizontal lines), which are topologically trivial and flat\cite{YYZhang2013}. In fact, they are so flat that their total measurement on the energy axis (sum of sub-band widths) even tends to zero in the thermodynamic limit\cite{YYZhang2012}. Therefore, these in-gap localized states will not affect the robust transports of edge states.

\subsection{B. Disorder Effect}

\begin{figure}[htbp]
\includegraphics*[width=0.6\textwidth]{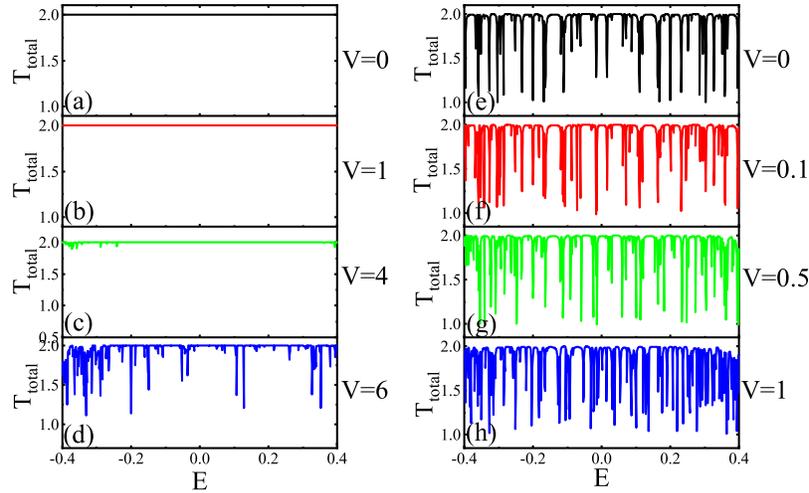}
\caption{(Color online) The total transmission as a function of Fermi energy $E$ at different disorder strength $\emph{V}$, for (a)-(d) $d_{2} = 2.5 $ and (e)-(h) $d_{2} = 1.2$. The rest model parameters are identical to figure  \ref{ConductanceEdgelocalityMarker}(a).}
\label{DisorderConductance}
\end{figure}

So far the results were calculated without any disorder, with the electronic properties as a consequence of nontrivial topology and incommensurate lattice. In this subsection, we investigate the role of disorder on this system. The effect of non-magnetic impurities is included by introducing a random potential term
\begin{equation}\label{RandomPotential}
\emph{V}_{\emph{I}}=\sum_{i\alpha}\emph{V}\emph{U}_\emph{i}\emph{c}_{\emph{i}\alpha}^{\dagger}\emph{c}_{\emph{i}\alpha}
\end{equation}
to the Hamiltonian, where $\emph{U}_\emph{i}$ are random numbers uniformly distributed in $\left(-0.5,0.5\right)$ and $\emph{V}$ is a single parameter to control the disorder strength. Total transmissions in the presence of different disorder strengths are presented in figure  \ref{DisorderConductance}, for $d_{2} = 2.5$ (left column) and for $d_{2} = 1.2$ (right column) respectively. At zero disorder, these two columns correspond to the central quantization plateau and transitional region respectively. In the former case, figure  \ref{DisorderConductance}(a)-(d), the quantized transmission plateau is unaffected by weak disorder until $V\sim 6$. In the latter case, figure  \ref{DisorderConductance}(e)-(h), more unquantized dips appear even
at weak disorder $V\sim 1$.

To characterize the stability and integrity of topological edge states in a more quantitative way,
we introduce the quantization index
\begin{equation}\label{PlateauIndex}
n_{\mathrm{Q}}=\frac{N_{\mathrm{Q}}}{N_{\mathrm{T}}},
\end{equation}
defined on a certain energy interval $[E_{1},E_{2}]$, where $N_{\mathrm{Q}}$ is the number of quantized data points whose total transmission $T_{\rm{total}}$ satisfies $|T_{\rm{total}}-2| < \xi$ with $\xi > 0$ a small tolerance error, and $N_{\mathrm{T}}$ is the total number of data points within this energy interval. Here we fix $\xi = 0.01$ and $[E_{1},E_{2}]=[-0.4,0.4]$.

\begin{figure}[htbp]
\includegraphics*[width=0.45\textwidth]{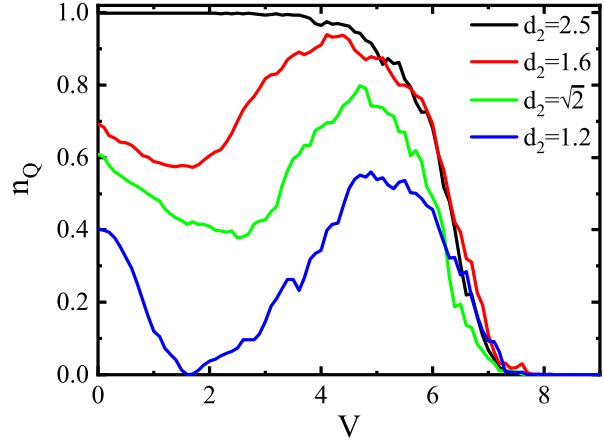}
\caption{(Color online) The quantization index $n_Q$ defined in equation (\ref{PlateauIndex}) as a function of disorder strength $\emph{V}$ at different $d_{2}$. $d_{2} = 1.2$ (blue solid lines), $d_{2} = \sqrt{2}$ (green solid lines),$d_{2} = 1.6$ (red solid lines), and $d_{2} = 2.5$ (black solid lines). The rest model parameters are identical to figure  \ref{ConductanceEdgelocalityMarker}(a).}
\label{PlateauIndex}
\end{figure}

\begin{figure}[htbp]
\includegraphics*[width=0.6\textwidth]{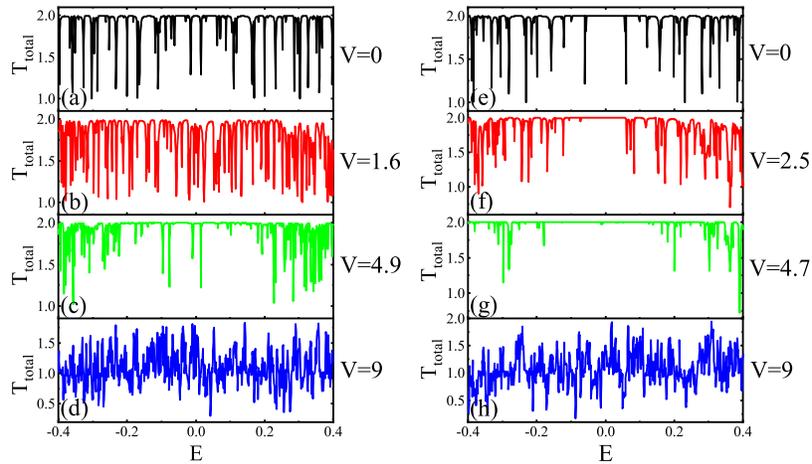}
\caption{(Color online) The total transmission as a function of Fermi energy $E$ at these specific transition disorder strength $V$ in figure  \ref{PlateauIndex}, for (a)-(d) $d_{2} = 1.2$ and (e)-(h) $d_{2} = \sqrt{2}$. The rest model parameters are identical to figure  \ref{ConductanceEdgelocalityMarker}(a).}
\label{TransitionConductance}
\end{figure}

The developments of the quantization index $n_{\mathrm{Q}}$ with increasing disorder strength $\emph{V}$ at different $d_{2}$ are presented in figure  \ref{PlateauIndex}. For the central quantization plateau, $d=2.5$ (black curve), it is perfectly intact as a whole with $n_{\mathrm{Q}}=1$ until
strong disorder $V\sim 3$. For the rest three cases, the energy region for calculating equation (\ref{PlateauIndex}) contains considerable portions of transitional regions in the presence of IGESs. As a result, as shown in figure  \ref{PlateauIndex}, all three curves of $n_{\mathrm{Q}}$ show a rapid response and a non-monotonic dependence on $V$. They first decrease at weak disorder, which suggests some of those quantization islands in the transitional region are destroyed. This originates from bulk and mutable IGESs, consistent with previous conclusions. Disorder can drive more IGESs into the bulk gap, or widen the existing IGESs. In either case, the interference with bulk IGES will destroy the quantized transport of edge states\cite{YYZhang2014}. Then, interestingly, they rise significantly at an intermediate disorder $V\in (2,5)$, indicating more quantized islands emerging. This can be directly confirmed from some snapshots of conductance at some $V$ shown in figure  \ref{TransitionConductance}, especially at $V=4.9$. Now, many IGESs have been localized and flattened by sufficiently strong disorder and they lose their contribution to quantum transports\cite{TAI,YYZhang2012,YYZhang2013}. Thus the topological edge states can manifest themselves on more Fermi energies. In other words, an intermediate disorder localizes IGESs and changes the picture from figure  \ref{FigIllustration} (a) to (b). At the very end, everything is localized in the strong disorder limit $V>7$, after the final touching of bulk bands and the annihilation of opposite Chern numbers\cite{YYZhang2012,TAIBeenakker}.

\subsection{C. Inter-layer Coupling}

\begin{figure}[htbp]
\includegraphics*[width=0.6\textwidth]{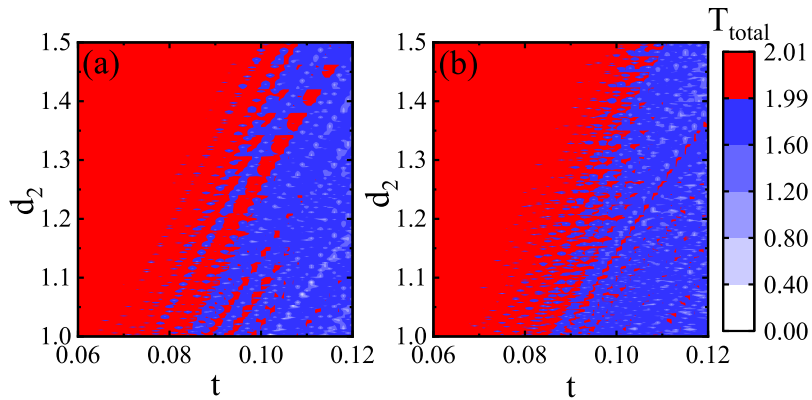}
\caption{(Color online) The total transmission on the inter-layer coupling strength-lattice constant plane with the central sample length and width (a) $L=W=50$, (b) $L=W=80$. The rest model parameters are: $A_{1} = 1$, $B_{1} = -1$, $D_{1} = 0$, $M_{1} = -1$, $A_{2} = 1$, $B_{2} = -1$, $D_{2} = 0$, $M_{2} = -1$, and $E = 1$.}
\label{CouplingConductance}
\end{figure}

Before the end, now we study the effects from varying the inter-layer coupling. The developments of the total transmission with the increasing of inter-layer coupling strength at different $d_2$ are illustrated in figure  \ref{CouplingConductance}. Let us start from a QAH phase at the inter-layer coupling $t=0$. Both panels of figure  \ref{CouplingConductance}(a) and (b) show that quantized transmission $T_{\rm{total}}=2$ persists at weak inter-layer coupling but is destroyed for large inter-layer coupling. Similar to the above case of varying $E$, there is also a transitional region outside the central region of perfect quantization, with inter-crossing structures of quantized and unquantized islands. Similarly, they can also be attributed to narrow bulk states as a result of lattice incommensurateness.

\section{V. Summary}

In this work, we studied the quantum transports of a bilayer QAH system. In the case of lattice match, we compared the transmissions from the tight-binding approach and the continuum Dirac model. Results from both methods have an excellent agreement especially in low-energy limit. We found that the intra-layer and inter-layer conductances show a complementarity since the sum of them is robustly quantized as 1 and the intra-layer transmission varies periodically with the central sample length. In this simple case, in the topological phase, the inter-layer coupling just leads to the resonant traversing between forward propagating waves in two layers.

In the case of lattice mismatch, the quantum transports are investigated by a full tight binding simulation. In the phase diagram on the $E-d_{2}$ plane, the central plateau of quantized conductance corresponds to the bulk gap which has not been influenced by the lattice mismatch.  On the other
hand, outside the central plateau of quantization, there is a transitional region with small and fractured islands of quantized and unquantized conductances penetrating each other. We identify this region as in-gap extended states penetrating with topological edge states. Details of this transitional region is sensitive to disorder, shape of the sample, and even the configuration of leads connected with it, due to the bulk and topologically trivial nature of these in-gap states. In phase diagram on the $E-t$ plane, the quantized conductance persists at weak inter-layer coupling and there is also a similar transitional region. Our results offer a comprehensive view of transport through the QAH bilayer with lattice mismatch.

\section{Acknowledgements}

This work was supported by National Natural Science
Foundation of China under Grant Nos. 11774336, 12104108, 61427901, and 12147112,
and the Starting Research Fund from
Guangzhou University under Grant Nos. RQ2020082 and 62104360.

\section{Appendix: Eigenvalues and Eigenfunctions of the Coupled Dirac Hamiltonian}
\setcounter{equation}{0}
\renewcommand{\theequation}{A\arabic{equation}}

The eigenvalues of the coupled Dirac Hamiltonian equation(\ref{DiracLikeHam}) are
\begin{equation}\label{Eigenvalues}
\begin{aligned}
&\varepsilon_{1}=\frac{1}{2}[\sqrt{4F^{2}+k_{x}^{2}(v_{1}-v_{2})^{2}}+k_{x}(v_{1}+v_{2})],\\
&\varepsilon_{2}=\frac{1}{2}[-\sqrt{4F^{2}+k_{x}^{2}(v_{1}-v_{2})^{2}}+k_{x}(v_{1}+v_{2})].
\end{aligned}
\end{equation}
Correspondingly, the normalized wave function takes the form
\begin{equation}\label{Eigenvectors}
\begin{aligned}
&\psi_{1}(x)=\left(
\begin{array}{c}
\frac{-\sqrt{4F^{2}+q_{1}^{2}(v_{1}-v_{2})^{2}}+q_{1}(v_{1}-v_{2})}{\sqrt{4F^{2}+[-\sqrt{4F^{2}+q_{1}^{2}(v_{1}-v_{2})^{2}}+q_{1}(v_{2}-v_{1})]^{2}}}\\
\frac{2F}{\sqrt{4F^{2}+[-\sqrt{4F^{2}+q_{1}^{2}(v_{1}-v_{2})^{2}}+q_{1}(v_{2}-v_{1})]^{2}}}             \\
  \end{array}
  \right)e^{iq_{1}x},\\
&\psi_{2}(x)=\left(
\begin{array}{c}
\frac{\sqrt{4F^{2}+q_{2}^{2}(v_{1}-v_{2})^{2}}+q_{2}(v_{1}-v_{2})}{\sqrt{4F^{2}+[-\sqrt{4F^{2}+q_{2}^{2}(v_{1}-v_{2})^{2}}+q_{2}(v_{1}-v_{2})]^{2}}}\\
\frac{2F}{\sqrt{4F^{2}+[\sqrt{4F^{2}+q_{2}^{2}(v_{1}-v_{2})^{2}}+q_{2}(v_{1}-v_{2})]^{2}}}             \\
  \end{array}
  \right)e^{iq_{2}x},
\end{aligned}
\end{equation}
where $q_{1}=\frac{\sqrt{E^{2}(v_{1}-v_{2})^{2}+4F^{2}v_{1}v_{2}}+E(v_{1}+v_{2})}{2v_{1}v_{2}}$ and $q_{2}=\frac{-\sqrt{E^{2}(v_{1}-v_{2})^{2}+4F^{2}v_{1}v_{2}}+E(v_{1}+v_{2})}{2v_{1}v_{2}}$.


\begin{thebibliography}{99}

\bibitem{Haldane} Haldane F D M 1988 {\it Phys. Rev. Lett.} {\bf 61} 2015
	
	\bibitem{QAH_CXLiu} Liu C X, Qi X L, Dai X, Fang Z and Zhang S C 2008 {\it Phys. Rev. Lett.} {\bf 101} 146802
	
	\bibitem{QAH_RYu} Yu R, Zhang W, Zhang H J, Zhang S C, Dai X and Fang Z 2010 {\it Science} {\bf 329} 61
	
	\bibitem{QAH_QZWang} Wang Q Z, Liu X, Zhang H J, Samarth N, Zhang S C and Liu C X 2014 {\it Phys. Rev.
	Lett.} {\bf 113} 147201
	
	\bibitem{QAH_HJZhang} Zhang H, Xu Y, Wang J, Chang K and Zhang S C 2014 {\it Phys. Rev.
	Lett.} {\bf 113} 216802
	
	\bibitem{QAH_SCWu} Wu S C, Shan G and Yan B 2014 {\it Phys. Rev. Lett.} {\bf 113} 256401

    \bibitem{QAHQSH1} Li B Y, Sun W L, Zou X R, Li X Y, Huang B B, Dai Y and Niu C W 2022 {\it New J. Phys.} accepted
	
	\bibitem{QAH_Exp1} Chang C Z, Zhang J, Feng X, Shen J, Zhang Z, Guo M,
	Li K, Ou Y, Wei P, Wang L L, Ji Z Q, Feng Y, Ji S,
	Chen X, Jia J, Dai  X, Fang Z, Zhang S C, He K, Wang Y, Lu L,
	Ma X C and Xue Q K 2013 {\it Science} {\bf 340} 167
	
	\bibitem{QAH_Exp2} Zhang J, Chang C Z, Tang P, Zhang Z, Feng X, Li K,
	Wang L L, Chen X, Liu C, Duan W, He K, Xue Q K, Ma X C and Wang Y, 2013 {\it Science} {\bf 339} 1582
	
	\bibitem{QAH_Exp3} Kou X, Guo S T, Fan Y, Pan L, Lang M, Jiang Y, Shao Q, Nie T,
	Murata K, Tang J, Wang Y, He L, Lee T K, Lee W L and Wang K L 2014 {\it Phys. Rev. Lett.} {\bf 113} 137201
	
	\bibitem{QAH_Exp4} Chang C Z, Zhao W, Kim D Y, Zhang H, Assaf B A, Heiman D,
	Zhang S C, Liu C, Chan M H W and Moodera J S 2015 {\it Nat. Mat.} {\bf 14} 473
	
	\bibitem{QAH_Exp5} Bestwick A J, Fox E J, Kou X, Pan L, Wang K L and Goldhaber-Gordon D 2015
	{\it Phys. Rev. Lett.} {\bf 114} 187201
	
	\bibitem{Bernevig2006} Bernevig A, Hughes T and Zhang S C 2006 {\it Science} {\bf 314} 1757
	
	\bibitem{GrapheneRMP} Castro Neto A H, Guinea F, Peres N M R, Novoselov K S and Geim A K 2009 {\it Rev. Mod. Phys.} {\bf 81} 109
	
	\bibitem{Bistritzer2011} Bistritzer R and MacDonald A H 2011 {\it Phys. Rev. B} {\bf 84} 035440
	
	\bibitem{YuanCao} Cao Y, Fatemi V, Demir A, Fang S, Tomarken S L, Luo J Y, Sanchez-Yamagishi J D, Watanabe K,
	Taniguchi T, Kaxiras E, Ashoori R C and Jarillo-Herrero P 2018 {\it Nature} {\bf 556} 80
	
	\bibitem{XiDai2019} Liu J, Ma Z, Gao J and Dai X 2019 {\it Phys. Rev. X} {\bf 9} 031021
	
	\bibitem{Twist2} Klug M J 2020 {\it New J. Phys.} {\bf 22} 073016

    \bibitem{Twistl} Li C F, Zhai W J, Li Y Q, Tang Y S, Zhang J H, Chen P Z, Zhou G Z, Cui X M,
	Lin L, Yan Z B, Huang X K, Jiang X P and Liu J M 2021 {\it New J. Phys.} {\bf 23} 083019
	
	\bibitem{FibonacciRMP} Jagannathan A 2021 {\it Rev. Mod. Phys.} {\bf 93} 045001
	
	\bibitem{JFanReview} Fan J and Huang H 2022 {\it Front. Phys.} {\bf 17} 13203
	
	\bibitem{MobilityEdge1D1} Li X, Li X and Das Sarma S 2017 {\it Phys. Rev. B} {\bf 96} 085119
	
	\bibitem{MobilityEdge1D2} Wang Y, Xia X, Zhang L, Yao H, Chen S, You J, Zhou Q and Liu X J 2020 {\it Phys. Rev. Lett.} {\bf 125} 196604
	
	\bibitem{YCZhang2022} Zhang Y C and Zhang Y Y 2022 {\it Phys. Rev. B} {\bf 105} 174206

    \bibitem{Update01} J. Wang, X.-J. Liu, X.-L Gao, and H. Hu 2016 {\it Phys. Rev. B} {\bf 93} 104504

    \bibitem{Update02} T. Liu, S. Cheng, H. Guo, and X.-L. Gao 2021 {\it Phys. Rev. B} {\bf 103} 104203

    \bibitem{Update03} S. Cheng, Y. Zhu, and X.-L. Gao 2022 {\it Symmetry} {\bf 14} 371

    \bibitem{Update04} S. Cheng and X.-L. Gao 2022 {\it Chin. Phys. B} {\bf 31} 017401

    \bibitem{Quasicrystal} Maniraj M, Lyu L, Mousavion S, Becker S, Emmerich S, Jungkenn D,
	Schlagel D L, Lograsso T A, Barman S R, Mathias S, Stadtm\"{u}ller B and Aeschlimann M 2020 {\it New J. Phys.} {\bf 22} 093056	
	\bibitem{HofstadterModel} Colandrea F D, D'Errico A, Maffei M, Price H M, Lewenstein M, Marrucci L, Cardano F,
	Dauphin A and Massignan P 2022 {\it New J. Phys.} {\bf 24} 013028
	
	\bibitem{Fabrication1} Ge J, Liu Y, Li J, Li H, Luo T, Wu Y, Xu Y and Wang J 2020 {\it Natl. Sci. Rev.} {\bf 7} 1280-1287
	
	\bibitem{Fabrication2} Zhao Y F, Zhang R, Mei R, Zhou L J, Yi H, Zhang Y Q,
	Yu J, Xiao R, Wang K, Samarth N, Chan M H W, Liu C X and Chang C Z 2020 {\it Nature} {\bf 588} 419
	
	\bibitem{Fabrication3} Algarni M, Tan C, Zheng G, Albarakati S, Zhu X, Partridge J, Zhu Y,
	Farrar L, Tian M, Zhou J, Wang X, Mao Z and Wang L 2022 {\it Phys. Rev. Lett.} {\bf 105} 155407
	
	\bibitem{FabricationOL1} Goldman N, Budich J C and Zoller P 2016 {\it Nat. Phys.} {\bf 12} 639
	
	\bibitem{FabricationOL2} Hartke T, Oreg B, Jia N and Zwierlein M 2020 {\it Phys. Rev. Lett.} {\bf 125} 113601
	
	\bibitem{FabricationOL3} Gall M, Wurz N, Samland J, Chan C F and K\"{o}hl M 2021 {\it Nature} {\bf 589} 40
	
	\bibitem{FabricationPhot} Khanikaev A B and Shvets G 2017 {\it Nat. Photonics} {\bf 11} 763
	
	\bibitem{QAH2010} Lu H Z, Shan W Y, Yao W, Niu Q and Shen S Q 2010 {\it Phys. Rev. B} {\bf 81} 115407

	\bibitem{InterlayerCoupling1} Slater J C and Koster G F 1954 {\it Phys. Rev. B} {\bf 94} 1498
	
	\bibitem{InterlayerCoupling2} Koshino M 2015 {\it New J. Phys.} {\bf 17} 015014
	
	\bibitem{Landauer1957} Landauer R 1957 {\it J. Res. Dev.} {\bf 1} 223--31
	
	\bibitem{Buettiker} B\"{u}ttiker M 1986 {\it Phys. Rev. Lett.} {\bf 57} 1761--4
	
	\bibitem{TransmissionRMP} Imry Y and Landauer R 1999 {\it Rev. Mod. Phys.} {\bf 71} S306--12

	\bibitem{Datta} Datta S 1995 {\it Electronic Transport in Mesoscopic Systems}
	(Cambridge: Canmbridge University Press)
	
	\bibitem{DHLee} Lee D H and Joannopoulos J D 1981 {\it Phys. Rev.} B {\bf 23} 4997--5004
	
	\bibitem{QuantumTransport} Khomyakov P A, Brocks G, Karpan V, Zwierzycki M and Kelly P J 2005 {\it Phys. Rev.} B {\bf 72} 035450
	
	 \bibitem{Jiang2009} Jiang H, Wang L, Sun Q F and Xie X C 2009 {\it Phys. Rev. B} {\bf 80} 165316
	
	\bibitem{YYZhang2013} Zhang Y Y and Shen S Q 2013 {\it Phys. Rev. B} {\bf 88} 195145
	
	\bibitem{NPR2021} Roy S, Mishra T, Tanatar B and Basu S 2021 {\it Phys. Rev. Lett.} {\bf 126} 106803
	
	\bibitem{NPR2020} Li X and Das Sarma S 2020 {\it Phys. Rev. B} {\bf 101} 064203
	
	\bibitem{NPR2017} Li X, Li X and Das Sarma S 2017 {\it Phys. Rev. B} {\bf 96} 085119
	
	\bibitem{HofstadterInsulators} Tran D Y, Dauphin A, Goldman N and Gaspard P 2015 {\it Phys. Rev. B} {\bf 91} 085125
	
	\bibitem{Dirac-likeHamiltonian1} Qi X L and Zhang S C 2011 {\it Rev. Mod. Phys.} {\bf 83} 1057
	
	\bibitem{Dirac-likeHamiltonian2} Zhou B, Lu H Z, Chu R L, Shen S Q and Niu Q 2008 {\it Phys. Rev. Lett.} {\bf 101} 246807
	
	\bibitem{BilayerHamiltonian1} Gonz\'{a}lez J W, Santos H, Pacheco M, Chico L and Brey L 2010 {\it Phys. Rev. B} {\bf 81} 195406
	
	\bibitem{BilayerHamiltonian2} Nilsson J, Castro Neto A H, Guinea F and Peres N M R 2007 {\it Phys. Rev. B} {\bf 76} 165416
	
	\bibitem{PerfectTransport1} Ando T, Nakanishi T and Saito R 1998 {\it J. Phys. Soc. Japan} {\bf 67} 2857-2862
	
	\bibitem{PerfectTransport2} McEuen P L, Bockrath M, Cobden D H, Yoon Y G and Louie S G 1999 {\it Phys. Rev. Lett.} {\bf 83} 5098-5101
	
	\bibitem{PerfectTransport3} K\"{o}nig M, Wiedmann S, Br\"{u}ne C, Roth A, Buhmann H, Molenkamp L W, Qi X L and Zhang S C 2007 {\it Science} {\bf 318} 766
	
	\bibitem{KleinParadox} Katsnelson M I, Novoselov K S and Geim A K 2006 {\it Nature Phys.} {\bf 2} 620-625
	
	\bibitem{GrapheneProperties} Castro Neto A H, Guinea F, Peres N M R, Novoselov K S and Geim A K 2009 {\it Rev. Mod. Phys. } {\bf 81} 109
	
	\bibitem{Current1} Ogawana T and Sakaguchi H 2021 {\it J. Phys. Soc. Jpn.} {\bf 90} 014002
	
	\bibitem{Current2} Dragoman D 2013 {\it J. Appl. Phys.} {\bf 113} 214312
	
	\bibitem{Current3} Peres N M R 2009 {\it J. Phys.: Condens. Matter} {\bf 21} 095501
	
	\bibitem{Current4} Snyman I, Tworzyd\l{}o J and Beenakker C W J 2008 {\it Phys. Rev. B} {\bf 78} 045118
	
	\bibitem{box-counting1} Guarneri I and Terraneo M 2001 {\it Phys. Rev. E} {\bf 65}, 015203(R)
	(2001).
	
	\bibitem{box-counting2} Veen E V, Yuan S J, Katsnelson M I, Polini M and Tomadin A 2016 {\it Phys. Rev. B} {\bf 93}, 115428
	
	\bibitem{TAI} Li J, Chu R L, Jain J K and Shen S Q 2009 {\it Phys. Rev. Lett.} {\bf 102} 136806
	
	\bibitem{YYZhang2014} Zhang Y Y, Shen M, An X T, Sun Q F, Xie X C, Chang K and Li S S 2014 {\it Phys. Rev. B} {\bf 90} 054205
	
	\bibitem{MeasurementPhaseTransition} Purkayastha A, Saha M and Agarwalla B K 2021 {\it Phys. Rev. Lett.} {\bf 127} 240601
	
	\bibitem{YYZhang2012} Zhang Y Y, Chu R L, Zhang F C and Shen S Q 2012 {\it Phys. Rev. B} {\bf 85} 035107
	
	\bibitem{TAIBeenakker} Groth C W, Wimmer M, Akhmerov A R, Tworzyd{\l}o J and Beenakker C W J 2009 {\it Phys. Rev. Lett.} {\bf 103} 196805
	
	
\end{thebibliography}
\end{document}